\documentclass{vgtc}                          

\ifpdf
  \pdfoutput=1\relax                   
  \usepackage{graphicx}                
  \DeclareGraphicsExtensions{.pdf,.png,.jpg,.jpeg} 
\else
  \usepackage{graphicx}                
  \DeclareGraphicsExtensions{.eps}     
\fi%

\graphicspath{{figures/}{pictures/}{images/}{./}} 

\usepackage{microtype}                 
\PassOptionsToPackage{warn}{textcomp}  
\usepackage{textcomp}                  
\usepackage{mathptmx}                  
\usepackage{times}                     
\usepackage{cite}                      
\usepackage{tabu}                      
\usepackage{booktabs}                  

\onlineid{0}

\vgtccategory{Research}

\vgtcinsertpkg

\title{Towards a Virtual Reality Home IoT Network Visualizer}

\author{Drew Johnston\thanks{e-mail: drewjj@rams.colostate.edu} %
\and Jarret Flack\thanks{e-mail:jflack@rams.colostate.edu} %
\and Indrakshi Ray\thanks{Indrakshi.Ray@colostate.edu} %
\and Francisco R. Ortega\thanks{e-mail:fortega@colostate.edu}}
\affiliation{\scriptsize Computer Science \\ Colorado State University}

\abstract{
We present an IoT home network visualizer that utilizes virtual reality (VR). This prototype demonstrates the potential that VR has to aid in the understanding of home IoT networks. This is particularly important due the  increased number of household devices now connected to the Internet. This prototype is able to function in a standard display or a VR headset. A prototype was developed to aid in the understanding of home IoT networks for homeowners.
} 

\CCScatlist{
  \CCScatTwelve{Human-centered computing}{Visu\-al\-iza\-tion}{Visu\-al\-iza\-tion techniques}{Treemaps};
  \CCScatTwelve{Human-centered computing}{Visu\-al\-iza\-tion}{Visualization design and evaluation methods}{}
}

\begin{document}


\firstsection{Introduction}
\maketitle

With the increased number of household devices now connected to the Internet, the personal data of the typical home-user will be increasingly tethered to the Internet of Things (IoT). Many users are sorely unaware of the personal data potentially leaked by these IoT devices in their home. Users are forced to trust that the devices on their network are behaving in a secure manner with little information regarding when, how, and to who their devices are communicating. We approached this problem with the assumption that the end-user does not have any formal understanding of computer networking. We therefore decided it best to work towards an intuitive visualization to give the typical home-user a better understanding of their own IoT device communications. We were able to identify two major questions while designing a visualization as such. Firstly, where does device network traffic get collected and processed? Secondly, with so much network traffic being produced, how can the user understand an effectively use the visualization without any background information in computer networking? We moved towards finding a solution to these questions through designing a 3D user interface (3DUI) to display IoT device network traffic. This system is able to function in a standard display or a virtual reality (VR) headset.  A prototype was developed with the purpose to show this proof of concept where networking students, researchers, and professional can benefit by using this type of visualization. Our main \textbf{contribution} is the ability to showcase a VR networking visualization system with an additional desktop version of the same. 

\section{Related Work}

Castelli et al. created a visualization dashboard system for end-users \cite{Castelli:2017:HMH:3025453.3025485}. This system was designed to visualize general network information for the home owner. The information provided included network speeds, security and other information related to home network. 

Ball et al. developed a system called VISUAL \cite{Ball:2004:HVN:1029208.1029217}. Their system included 2D graphics to display information. There focus was on the visualization of network traffic in homes that with the assumption that they are vulnerable to cyber-attacks. This has some similarities with our system with the difference that the information is displayed using 2D user interfaces (leading to a notably cluttered display). 

Some of our design decisions were influenced by ideas presented by Le Mal{\'e}cot et al \cite{LeMalecot:2006:ICV:1179576.1179600}. They found that visualizing network data can be extremely confusing and cluttered when display in 2D user interfaces. 3DUIs provide a more intuitive understanding of the network packets and nodes. Additional work in this area can be found in \cite{Poole:2008:MME:1394445.1394494,7298088}. 



\section{Virtual Reality Networking Visualization}

\begin{figure}[!tb]
\centering 
\centering 
\includegraphics[width=\columnwidth]{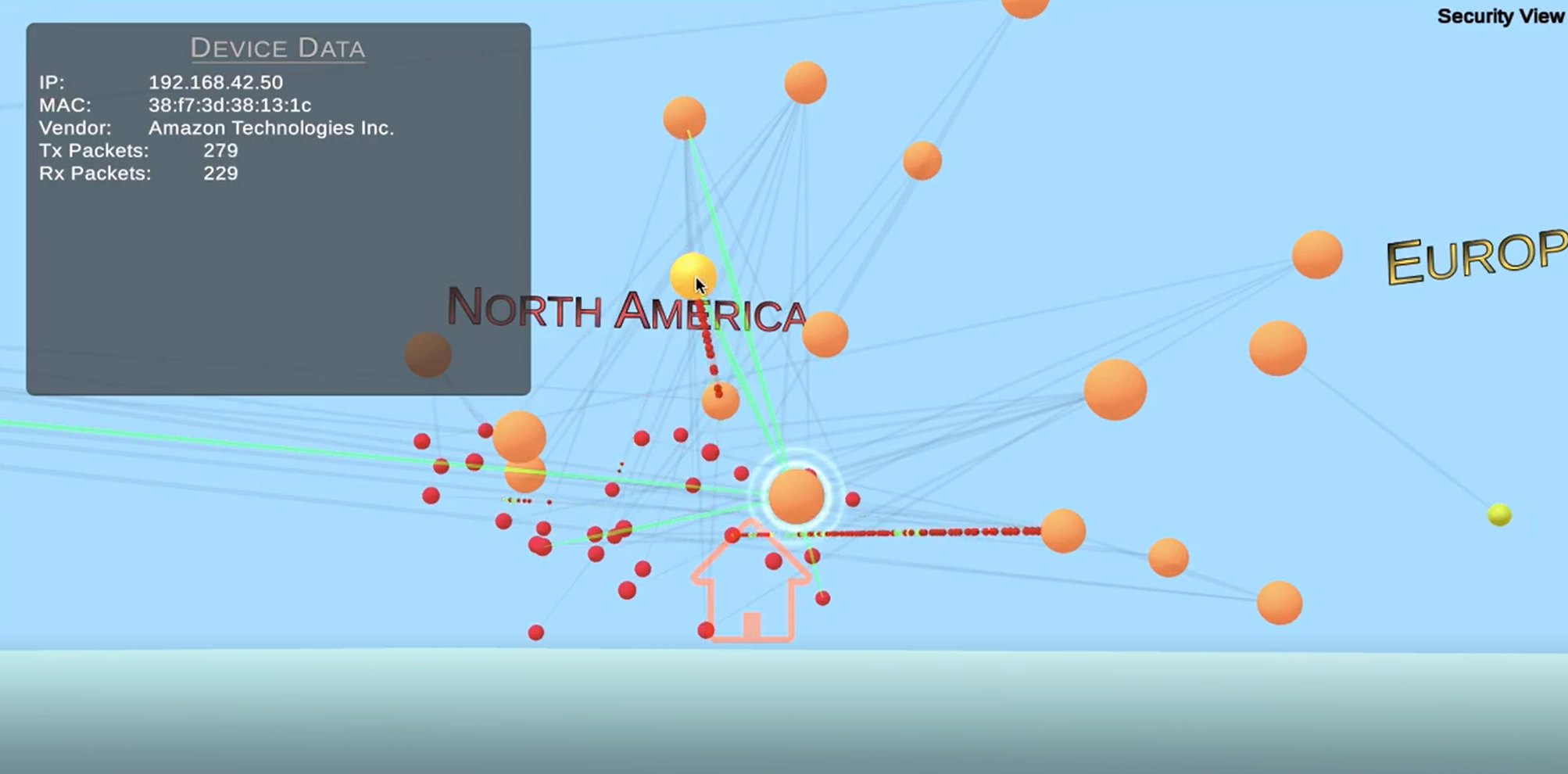}
\caption{3DUI Visualization}
\label{fig:3dui1}
\end{figure}

Our prototype system is comprised of two main components. The first is a back-end server that handles the collection, parsing, and querying of network traffic. The second component is a front-end program that produces a dynamic visualization of the network traffic based on data streamed from the back-end server.

Currently, the collection of network traffic is facilitated by the back-end server. The server communicates with a machine utilizing host access point daemon (hostapd)  to act as an access point for several real IoT devices. When the server desires to collect network traffic data from the IoT devices, it will remotely run a script on the access point and pull the data back to the server. This collected data is then processed accordingly and placed into a database to be queried by the user from the front-end. When the back-end server processes collected data, it keeps a cache of identified device specific information to be leveraged on the front-end, such as vendor identification or location. It should be noted that every record placed into the database is a list of features that describes a single network packet that passed through the access point, device specific information included.

The packets collected can be moving out or into the local area network or also packets moving between devices within the local area network. Therefore, the system is able to examine any traffic that leaves, enters, or exists within the local area network (LAN). The back-end server is able to communicate with any number of front-end clients, whenever a client requests data. We have developed a custom method of streaming data to front-end clients that allows for responsive queries regardless of the amount of data returned. For every data request, the back-end server dynamically builds serializable objects to represent devices, connections, and individual packets which are then streamed to the front-end.

The visualization, as shown in Figure \ref{fig:3dui1}, provides an interactive user experience that allows exploration of nodes and individual network packets, setting our work apart from existing networking visualization techniques. The ability to have an interactive VR experience, users can appreciate networking information as never seen before. For example, the ability to understand that a HTTP message is traveling from multiple routes while using the virtual environment (VE) provided, gives a sense of immersion while understanding networking.

Users are able to customize a query in the desktop by using traditional keyboard or mouse, or using the HTC Vive controllers in the VR experience. Currently, the data is stored in a back-end server database. The back-end server streams the requested traffic back to the VE. The user experiences the packets as they arrive from the back-end in real-time (note that the data is stored in the server but the streaming is real-time). Each device and server is represented as a selectable spherical node, nodes are clustered by location. The devices in a user’s home are clustered together around an orange house icon. In addition, every remote device or server they may connect to is clustered near the name of the continent they are located in (e.g., North America). Users can observed how hosts interact with their local area network. Packets are represented by small spheres that travel between hosts real-time, the color of the sphere depends on the protocols used for that packet and the view selected. Lines will also be drawn between nodes when they initially send data to each other to represent connections. The user can then select a local host or remote host. Once selected, the system will provide network analytics as well as highlight all of the connections of the given device. This visualization is based on the historical captured data of the home network so it is an accurate and viewable representation of that home network.

As already mentioned, we have two prototypes. A desktop 3DUI version and a virtual reality version. While the desktop 3DUI version provides interesting information, it is the virtual reality experience that allows the immersion of the user. While no formal testing was conducted at this stage, people have interacted with the system. Even for people with no networking experience, found the VR system useful to understand packets and internet traffic. In particular, networking professionals from [Company1] and [Company2] were interested in the potential that this project has. However, at the current state of this project, the desktop version was easier for interaction. Yet, the VR environment interaction can be improved and we do think this is the main path for future work. 

The system was developed in Unity Game Engine using an HTC Vive. The desktop version, also developed in Unity, has been tested in Mac OSX and Windows. The server is linux-based and the back-end system was developed in Python. 

\begin{figure}[!tb]
\centering 
\centering 
\includegraphics[width=1.0\columnwidth]{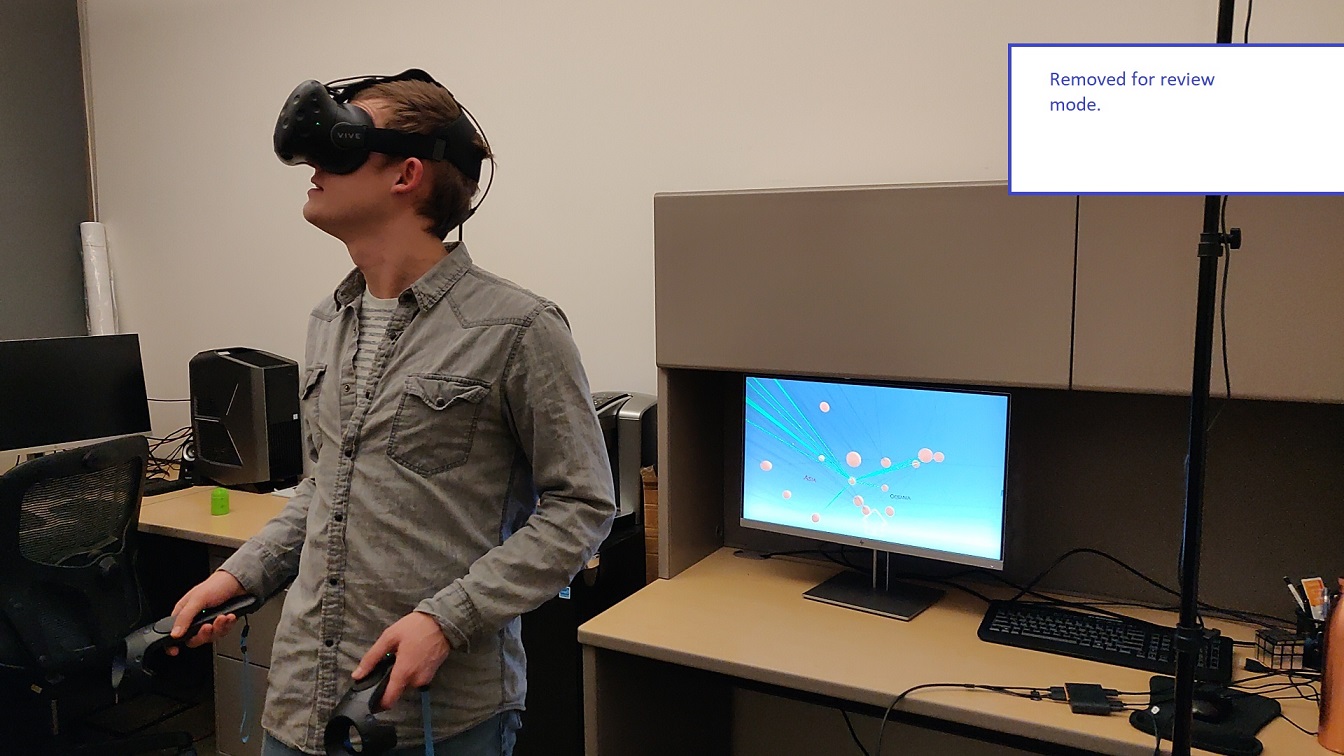}
\caption{Visualization with HTC Vive}
\label{fig:vr}
\end{figure}

\section{Future Work}
To approach a more optimal solution, there are multiple improvements and research questions that need to be addressed. The current prototype only supports querying network traffic that had been previously collected; therefore, the need to handle real-time data (beyond just streaming real-time) is critical. In addition, some possible other avenues for future work may include real-time solutions to preform data analytics and anomaly detection. Furthermore, due to the volume of network traffic produced by these devices, one research question that is important to answer is how to aggregated on the back-end while reducing the computational load in the VR system. Another important aspect of future work includes the the ability for users to prevent devices from communicating using SNORT rules dynamically applied to the network. While this can be easily accomplished in a desktop environment, it is an open question, what technique would work best for interacting with complex rules using VR. Finally, the most immediate step is to find networking students and/or professionals to conduct a user study and compare the gains between this prototype using VR, an existing 2D visualization system, and no visualization system. This may provide comparison measurements to see the advantage, if any, of using a VR networking visualization system. 

\section{Conclusion}

As the home environment becomes increasingly connected to the Internet, there is a need to increase the general public's understanding of how their IoT devices interact with entities outside and within their local area network. We believe that the visualization system presented will make a difference in the understanding of these IoT networks.


\bibliographystyle{abbrv-doi}

\bibliography{template}
\end{document}